\documentclass{aa}  
\usepackage[normalem]{ulem}
\usepackage{graphicx}
\usepackage{xcolor}
\usepackage{txfonts}
\usepackage{longtable}

\begin{document}

\title{Resonant absorption: \\ Transformation of compressive motions into vortical motions}
\titlerunning{Transformation of compressive motions into vortical motions}

\author{M. Goossens
        \inst{1}
        \and
        I. Arregui 
        \inst{2,3}\
        \and
        R. Soler 
        \inst{4,5}\
        \and
        T. Van Doorsselaere
        \inst{1}
}
\institute{Centre for mathematical Plasma Astrophysics, KU Leuven, Celestijnenlaan 200B bus 2400, B-3001 Leuven, Belgium
        \and Instituto de Astrof\'{\i}sica de Canarias, V\'{\i}a L\'actea s/n, E-38205 La Laguna, Tenerife, Spain\
        \and
        Departamento de Astrof\'{\i}sica Universidad de La Laguna, E-38206 La Laguna, Tenerife, Spain\
        \and
        Departament de F\'{\i}sica, Universitat de les Illes Balears, E-07122 Palma de Mallorca, Spain\
        \and
        Institut d$'$Aplicacions Computacionals de Codi Comunitari (IAC$^3$), Universitat de les Illes Balears, E-07122 Palma de Mallorca, Spain\\
        \email{marcel.goossens@kuleuven.be}
}

\abstract{
This paper investigates the changes in spatial properties when magnetohydrodynamic (MHD) waves undergo resonant damping in the Alfv\'en continuum.  The analysis is carried out for a 1D cylindrical pressure-less plasma with a straight magnetic field.  The effect of the damping on the spatial wave variables is determined by using complex frequencies that arise as a result of the resonant damping.  Compression and  vorticity are used to characterise the spatial evolution of the MHD wave. The most striking result is the huge spatial variation in the vorticity component parallel to the magnetic field. Parallel vorticity vanishes in the uniform part of the equilibrium. However,  when the MHD wave moves into the non-uniform part, parallel vorticity explodes to values that are orders of magnitude higher than those attained by the transverse  components in planes normal to the straight magnetic field. In the non-uniform part of the equilibrium plasma, the MHD wave is controlled by parallel vorticity and resembles an Alfv\'en wave, with the unfamiliar property that it has pressure variations even in the linear regime.}


\maketitle

\section{Introduction}

In a recent paper, \cite{goossens19} have studied the properties of  magnetohydrodynamic (MHD) waves in non-uniform plasmas. They pointed out that in non-uniform plasmas, MHD waves behave differently from their counterparts in uniform plasmas of infinite extent.  In the latter case, the MHD waves can be separated into Alfv\'en waves and magneto-acoustic waves.  The Alfv\'en waves propagate vorticity and are incompressible. In addition, they have no parallel displacement component. The magneto-acoustic waves are compressible and in general  have a parallel component of displacement, but do not propagate parallel vorticity. Compression, parallel vorticity, and parallel displacement are the characteristic quantities. In a uniform plasma of infinite extent, compression and parallel displacement on one hand and parallel vorticity on the other hand are mutually exclusive. In a pressure-less plasma, the parallel displacement is zero because the Lorentz force has no component along the magnetic field.  Hence, in a pressure-less plasma, the waves have only two characteristic quantities: compression and parallel vorticity.  The distinction between the waves is then based on compression or parallel vorticity.

The situation is different in a non-uniform plasma, as was pointed out by \cite{goossens19}.  The clear division between Alfv\'en waves and magneto-acoustic waves is no longer present. The MHD waves have mixed properties. The general rule is that MHD waves propagate both parallel vorticity, as in classic Alfv\'en waves, and compression, as in classic magneto-acoustic waves.  The present paper focusses on the properties of MHD waves that undergo resonant absorption. Here we concentrate on resonant absorption in the Alfv\'en continuum. In order to keep the mathematical analysis as simple as possible while still retaining the essential physics, we consider a straight magnetic field, and in addition, we assume that the plasma is pressure-less. This assumption removes the slow magneto-acoustic part of the spectrum and resonant absorption in the slow continuum.  

\cite{goossens19}  also studied what happens with compression and vorticity for frequencies in the slow and Alfv\'en continuum. The  analysis was restricted to the driven problem of stationary waves with real frequencies.  The authors determined the dominant singularities in the ideal MHD solutions and the dominant dynamics  for stationary waves.  For another study in which resonant absorption in both Alfv\'en and slow continua are considered, see \cite{soler09}. Studies on resonant absorption have mainly focussed on the components of the displacement, the amount of absorbed energy, and the damping rate. Analytical solutions for the components of the Lagrangian displacement in the dissipative layer have for example been derived by  \cite{sakurai91} and \cite{goossens95}  for  resonant MHD waves in ideal and dissipative stationary MHD.   \cite{ruderman95} studied non-stationary incompressible resonant MHD waves in non-ideal MHD for a planar equilibrium. \cite{tirry96}  studied non-stationary resonant MHD waves for cylindrical plasmas in visco-resistive  MHD.  \cite{soler13} studied non-stationary MHD waves for cylindrical plasmas both in resistive MHD and ideal MHD. Their mathematical scheme for non-stationary ideal MHD followed the scheme devised by \cite{hollweg90b} for planar plasmas. Little or no attention has been given to the change in the spatial behaviour of fundamental quantities such as compression and vorticity.  \cite{goossens12} were the first to point out that the fundamental radial mode of kink waves propagates parallel vorticity in the non-uniform part of the loop required for resonant absorption to operate.  
In the present investigation, we focus on the eigenvalue problem and try to understand what happens when the wave is actually damped in non-stationary MHD. We take the frequency to be complex and relate the spatial behaviour to the damping properties of the MHD wave. \cite{soler13} concentrated on the components of Lagrangian displacement and the Eulerian perturbation of total pressure. \cite{goossens12} took the existence of a parallel vorticity component as the base for the physical interpretation of the fundamental radial mode of kink waves in terms of surface Alfv\'en waves.  Our analytic investigation first presents the components of vorticity in the case of non-stationary ideal MHD close to the resonant position. Then, the semi-analytic approach of  \cite{soler13} in non-stationary MHD is used to verify the predictions based on approximate analytic theory. The aim is to obtain a simple understanding of how the fundamental quantities compression and vorticity are affected by the non-stationary behaviour of the resonantly damped wave.
 
\section{Resonant absorption} 

This section collects results on resonant absorption that are used in our discussion of the spatial solutions of the MHD waves that undergo resonant damping. Resonant absorption has a long history in fusion plasma physics, space plasma physics, solar physics, and astrophysics.  A characterisation was given by  \cite{parker91}, who noted that resonant absorption in the Alfv\'en continuum is to be expected when  a wave with a phase velocity 
$\omega / k$ spans a region in which the variation of the Alfv\'en velocity $v_{\rm A}$ across the region provides the  resonance condition  $\omega / k = v_{\rm A}$. Non-uniformity is key  to the process.  The Alfv\'en velocity $v_{\rm A}$  and  Alfv\'en frequency $\omega_{\rm A}$ depend on position.  The resonance occurs at the position $r_{\rm A}$ , where the frequency of the wave $\omega$ is equal to the local Alfv\'en frequency  $\omega = \omega_ {\rm A}(r_{\rm A})$.  The local Alfv\'en frequency $\omega_{\rm A}$ is assumed to be an analytic function.

Since 2002 \citep{ruderman02,goossens02a}, resonant absorption of kink waves is a popular and plausible mechanism for explaining the rapid damping of standing and propagating MHD waves in coronal loops \citep[see e.g.][for a discussion]{montessolis17}. The simple model invokes MHD waves superimposed on a cylindrical plasma column in static  equilibrium. Cylindrical coordinates $(r, \varphi, z)$ are used. The inhomogeneity necessary for resonant absorption to operate is usually provided by the equilibrium density $\rho_{\rm 0}(r)$ that varies  from  $\rho_{\rm i}$ to $\rho_{\rm e}$ in the interval $[R - l/2, \;\;R + l/2]$.  The density $\rho_{\rm 0}$ is constant in the internal and the external parts of the loop with values $\rho_{\rm i}$ and $\rho_{\rm e,}$

\[ \rho_{\rm 0} = \begin{cases} \rho_{\rm i} &  \mbox{ for}\;\;  0  \leq r \leq R -l/2,   \\
\rho_{\rm e} &   \mbox{ for}\;\; R + l/2 \leq r < + \infty. \end{cases} \]

In the non-uniform transitional layer of thickness $l,$ the equilibrium density varies continuously from its internal value $\rho_{\rm i}$ to its external value $\rho_{\rm e}$ .

The equilibrium magnetic field is assumed to be axial and constant 
 $\vec{B}_0$ = $B_0 \vec{1}_z$,  with $\vec{1}_z$ the unit vector in the $z$- direction. The temporal dependence and the spatial dependence on the ignorable $(\varphi, z)$- coordinates are given by  $  \exp(- i \omega t)$ and $\exp (i (m\varphi + k_{\rm z}z))$,  with $m$ and $k_{\rm z}$  the  azimuthal and axial wave numbers, respectively. The local Alfv\'en frequency $\omega_{\rm A}$ and the Alfv\'en velocity $v_{\rm A}$ are defined as 

\begin{equation}
 \omega_{\rm A}^2  = \frac{ (\vec{k} \cdot
\vec{B})^2}{ \mu \rho_{\rm 0}} \;=\;k_{\rm z}^2\;v_{\rm A}^2 = k_{\parallel}  v_{\rm A}^2\mbox{\hspace{0.5cm}} \mbox{\rm and} \mbox{\hspace{0.5cm}}   v_{\rm A}^2 = \frac{B_{\rm 0}^2}{ \mu  \rho_0}.
\label{AlfvenFreVA}
\end{equation}

The fundamental radial mode of  kink ($m$=$1$)  waves has its frequency in the Alfv\'en continuum and is always  resonantly damped. Here we consider standing waves, which means that the axial wave number $k_{\rm z}$ is real and the  frequency $\omega$ is complex. Hence

\begin{eqnarray}
\omega & =&  \omega_{\rm R} + i \gamma,  \nonumber \\
\exp(- i \omega t)&  = & \exp(- i \omega_{\rm R} t)\exp(\gamma t) = \exp(- i \omega_{\rm R} t)\exp(- t/ \tau_{\rm D})\nonumber\\
\label{Dampedexp}
.\end{eqnarray}

The  ${\rm Period}$ =  $2\pi / \omega_{\rm R}$,  $\gamma$ is the decrement, and  $\tau_{\rm D} = 1/\mid \gamma \mid$ the damping time. The  variation in density is confined to a layer of thickness $l$ and has a steepness $\alpha$. The steepness $\alpha$ is related to the factor $F$ introduced by \cite{arregui07,arregui08b} and has been used many times before, for example, by \cite{goossens08a}, \cite{arregui08a}, and \cite{soler09}  as
\begin{equation}
F=\alpha\frac{4}{\pi^2}.
\end{equation}
When we aim to arrive at analytical expressions for the period and damping time, we can use the thin tube and  thin boundary (TTTB)  approximation so that  $k_{\rm z} R  \ll 1$, $l/R  \ll 1$. Analytical expressions for the damping time $\tau_{\rm D}$  were derived by \cite{hollweg90a}, \cite{goossens92}, \cite{ruderman02}, \cite{vandoorsselaere04}, \cite{soler13}, and \cite{arregui19}. We use the expression given by \cite{arregui19},

\begin{equation}
 \frac{\tau_{\rm D}}{\mbox{Period}} = \frac{ 1}{\mid m \mid} \;\frac{4}{\pi^2}\; \frac{\alpha}{l/R} 
\;\frac{\rho_{\rm i} + \rho_{\rm e}}{\rho_{\rm i} -\rho_{\rm e}}. 
\label{TauAG}
\end{equation}

In this equation, the quantities $l/R$ and $\alpha$ arise from the adoption of variation in the equilibrium density in a non-uniform layer of thickness $l$ and with steepness $\alpha$ so that the derivative of the density at the resonant position $r_{\rm A}$ is given by

\begin{equation}
\frac{d \rho_0}{d r}]_{r = \operatorname{Re}(r_{\rm  A})} = - \alpha \;\frac{\rho_{\rm i} - \rho_{\rm e}}{l}.
\label{DerivativeD}
\end{equation}
We first recall that $\omega$ is complex, with an imaginary part $\gamma,$ and strictly speaking, the position $r_{\rm A}$ is in the complex plane.  The thin boundary (TB) approximation assumes that the absolute value of $\gamma$ is low in comparison with $\omega_{\rm R}$ and that $\operatorname{Im}(r_{\rm A})$ is neglected. In what follows, $\operatorname{Im}(r_{\rm A})$ is neglected unless otherwise stated.

The steepness $\alpha $ is 1 for a linear variation of density and $\pi /2$ for a sinusoidal variation. For kink waves, $m=1$, but any non-axisymmetric wave ($m\neq0$) with its frequency in the Alfv\'en continuum is resonantly damped. \cite{soler17} studied the resonant damping of fluting modes ($|m| \ge 2$) and showed that these modes can be heavily damped ($\tau_{\rm D}/{\rm Period} \ll 1$).
We note also that the non-uniform layers are often thick and the condition $l/R \ll 1$ is not satisfied \citep[see e.g.][]
{goossens02a,arregui07,goossens08a,pascoe19}. The errors associated with the use of the TTTB approximation beyond its theoretical range of applicability have been estimated by \cite{vandoorsselaere04} 
and \cite{soler14}.

\section{MHD waves with mixed properties for a straight field}

We recall from \cite{goossens19} that the mixed properties arise because the MHD 
equations are coupled. The coupling of the MHD equations is controlled by the  coupling functions  $ C_{\rm A}$ and $C_{\rm S}$  , which were first introduced by \cite{sakurai91}. 
For a straight field  $\vec{B_{\rm 0}}= B_{\rm 0}(r) \vec{1}_{\rm z}$, they take the simple form 

\begin{equation}
C_{\rm A} =  g_{\rm {B}} \;P' = \frac{m}{r} B_{\rm 0} \;P', \;\; C_{\rm S} = P',
\label{CASStraightF}
\end{equation}
with $P'$ the Eulerian perturbation of total pressure. This means that the equations are  coupled because of $P'$. This was already known by \cite{hasegawa82}. They noted that \begin{quote}The basic characteristic of the ideal Alfv\'en wave is that the total pressure in the fluid remains constant during the passage of the wave as a consequence of the incompressibility condition. For  inhomogeneous medium, however, the total pressure, in general, couples with the dynamics of the motion, and the assumption of neglect of pressure perturbations becomes invalid. \end{quote}

We now concentrate on  compression and parallel vorticity. We also take  non-axisymmetric waves $m \neq 0$ because for $m=0$, $C_A = 0,$ and there is no coupling of equations, no mixed properties, and no resonant damping. Because we study non-stationary MHD waves, the frequency $\omega$ that appears in the following equations is a complex quantity, as defined in Eq.~(\ref{Dampedexp}).  For a  pressure-less plasma, the velocity of sound  $v_{\rm S} = 0$ and  there are no parallel motions $\xi_{\rm z} = \xi_{\parallel} = 0$. We take $v_{\rm S}^2 = 0$ in equations (45) of \cite{goossens19} and find  for the components of the displacement $\vec{\xi}$ and for compression $\nabla\cdot \vec{\xi}$ 

\begin{eqnarray}
\xi_r & = & \frac{1}{\rho_{\rm 0} (\omega^2 - \omega_{\rm A}^2)} \frac{d P'}{dr}, \nonumber \\
&& \nonumber \\ 
\xi_{\perp} & = & \xi_{\varphi} = \frac{ i m/r}{\rho_{\rm 0} (\omega^2 - \omega_{\rm A}^2)} P',  \nonumber \\
&& \nonumber  \\
\xi_{\parallel} & = & \xi_{\rm z} = 0 ,\nonumber  \\
&& \nonumber \\
\nabla \cdot \vec{\xi} & = &  \frac{- P' }  { \rho_{\rm 0}  v_{\rm A}^2} = \frac{- P' }  { B_{\rm 0}^2 /\mu_0}.
\label{EqStrFieldA}
\end{eqnarray}

Similarly, we take  $v_{\rm S}^2 = 0$ in equations (53) of \cite{goossens19} and find  for the components of  vorticity $\nabla \times \vec{\xi}$ 

\begin{eqnarray}
(\nabla \times \xi)_{\rm r}  & = & k_{\rm z}  \frac{m}{r}  \frac{1}{\rho_{\rm 0} (\omega^2 - \omega_{\rm A}^2)}P'. \nonumber \\
&& \nonumber \\
(\nabla \times \vec{\xi})_{\perp} & = &   (\nabla \times \vec{\xi})_{\varphi} = 
i k_{\rm z}   \frac{1}{\rho_{\rm 0} (\omega^2 - \omega_{\rm A}^2)}  \frac{d P'}{dr}, \nonumber \\
&& \nonumber \\
(\nabla \times \vec{\xi})_{\parallel} & = & (\nabla \times \xi)_z \nonumber\\
&=&-  i \frac{m}{r} \frac{\displaystyle 1} {\displaystyle \left \{\rho_{\rm 0} (\omega^2 - \omega_{\rm A}^2) \right \} ^2}
\frac{\displaystyle d}{\displaystyle dr} \left \{\rho_{\rm 0} (\omega^2 - \omega_{\rm A}^2)\right \}  P'.
\label{EqStrFieldB}
\end{eqnarray}

Equations~(\ref{EqStrFieldA}) and (\ref{EqStrFieldB})  clearly show that  $P'$ plays the role of the coupling function.  The transverse components of vorticity $(\nabla \times \xi)_{\rm r} $ and $(\nabla \times \vec{\xi})_{\varphi}$ are always non-zero. The parallel component $(\nabla \times \vec{\xi})_{\parallel} = (\nabla \times \xi)_z$ is non-zero when

\begin{equation}
\frac{\displaystyle d}{\displaystyle dr} \left \{\rho_{\rm 0} (\omega^2 - \omega_{\rm A}^2)\right \} \neq 0.
\label{NonUniformOmegaA}
\end{equation}

All wave variables are non-zero in a non-uniform plasma.   Except that here $\xi_{\parallel} = 0$  because of the assumption $v_{\rm S}^2 = 0$.  There are  no pure magneto-acoustic waves and no pure Alfv\'en waves  in a non-uniform plasma.  The MHD waves have mixed properties, and they  also behave differently in different parts of the plasma because of the inhomogeneity of the plasma, as already emphasised, for example, by \cite{goossens08b}, \cite{goossens02a}, \cite{goossens06}, \cite{goossens11}, \cite{goossens12}, and \cite{goossens19}.

We note  that for a straight constant field $\vec{B}_0 = B_{\rm 0} \vec{1}_z $ , the spatial variation of $\omega_{\rm A}^2$ is solely due to the spatial variation of the equilibrium density $\rho_{\rm 0}$,  hence 

\begin{equation}
\frac{\displaystyle d\rho_{\rm 0}}{\displaystyle dr} \neq 0
\label{NonDensity}
\end{equation}
is the important quantity for the resonant behaviour.  The expression for the parallel component of vorticity then simplifies to 

\begin{equation}
(\nabla \times \vec{\xi})_{\parallel} = (\nabla \times \xi)_z = -  i \frac{m}{r} \frac{\displaystyle  \omega^2 } {\displaystyle \left \{\rho_{\rm 0} (\omega^2 - \omega_{\rm A}^2) \right \} ^2}\;
\frac{\displaystyle d  \rho_{\rm 0}}{\displaystyle dr} \; P'.
\label{vorticityZ1}
\end{equation}

In order to make clear, as was done by \cite{goossens19} in their equations (62, 63, and 64),  that parallel vorticity and compression go together in a non-uniform plasma, the previous Eq.~(\ref{vorticityZ1}) can be rewritten as
 
\begin{equation}
  (\nabla \times \vec{\xi})_{\parallel} = (\nabla \times \xi)_z =  i \frac{m}{r} \frac{\displaystyle  \omega^2 } {\displaystyle \left \{\rho_{\rm 0} (\omega^2 - \omega_{\rm A}^2) \right \} ^2}\;
\frac{\displaystyle d  \rho_0}{\displaystyle dr} \; \frac{ B_{\rm 0}^2}{  \mu_{\rm 0}} \;\;\nabla \cdot \vec{\xi}.
\label{vorticityZ2}
\end{equation}
We note that 
$$
\rho_0 v_A^2  =  \frac{B_0^2}{\mu_0} = \mbox{constant}, \;\;
\rho_0  \omega_A^2 = k_z ^2 \frac{B_0^2}{\mu_0} = \mbox{constant} . 
$$

\section{Analysis close to the ideal resonant point}

This section investigates the behaviour of the wave variables for a standing damped resonant wave close to the ideal resonant point $r_{\rm A}$.  For a resonantly damped standing wave, Eq.~(\ref{Dampedexp}) reads 

\begin{eqnarray}
\omega  & = & \omega_{\rm R} + i  \gamma,  \;\;\gamma = \omega_{\rm I}, \;\ \omega_{\rm R} \approx  \omega_{\rm A}(r_{\rm A}), \nonumber \\
\frac{\mid \gamma \mid}{\omega_{\rm R}} & = &  \frac{1}{2 \pi} \frac{\mbox{Period}}{\tau_{\rm D}}.
\label{DampedRW}
\end{eqnarray}

An immediate consequence of the fact that the frequency, $\omega$, is complex in non-stationary ideal MHD is that the resonant point, $r_{\rm A}$, is also a complex quantity. The higher the damping rate, $\gamma$, the larger the imaginary part of $r_{\rm A}$. The following analytic investigation assumes weak damping, so that the imaginary part of $r_{\rm A}$ can be ignored. We therefore treat $r_{\rm A}$ as a real quantity as in stationary MHD. However, we are aware that this is just an approximation. In Section 5 we show that we need to consider $r_{\rm A}$ to be complex in order to understand the behaviour of the perturbations when wave damping is not weak.

It is instructive to summarise the main results on the behaviour of resonant MHD waves in ideal MHD. The  fundamental conservation law for resonant Alfv\'en waves for a straight field was obtained by \cite{sakurai91},

\begin{equation}
 P' = \mbox{constant}, \;\; [P'] = 0.
\label{CAstraightconstant}
\end{equation}
This conservation law was later confirmed in dissipative stationary MHD by Goossens et al. (1995) for cylindrical plasmas, in dissipative non-stationary MHD  for incompressible plasmas by Ruderman et al. (1995) for  planar geometry, and by \cite{tirry96} for  compressible visco-resistive MHD in cylindrical geometry.  \cite{soler13} showed that in non-stationary ideal MHD, $P'$ displays a logarithmic jump at the resonant point. The jump is proportional to the imaginary part of $r_{\rm A}$, so that the conservation law of $P'$ = ${\rm constant}$ is approximately valid if the damping is only weak 
(see Section 5).

In view of the observed periods and damping  times, it is an accurate approximation to use 

 \begin{equation}
\frac{\mid \gamma \mid}{\omega_{\rm R}} \ll 1.
\label{weakdamping}
\end{equation}

In the first instance we concentrate on the parallel component of vorticity. The analysis in \cite{goossens19} has shown that in ideal  MHD, the dominant dynamics resides in the perpendicular component of the displacement and the parallel component of vorticity in ideal MHD.  We use the following approximate results by retaining the dominant terms, that is, the lowest order terms in $\mid \gamma \mid / \omega_{\rm R}$:

\begin{eqnarray}
\omega^2 \frac{d \rho_{\rm 0}}{dr}  & = & \omega_{\rm R}^2 (1 + i \frac{\gamma}{\omega_{\rm R}})^2\; \frac{d \rho_{\rm 0}}{dr}  \; \approx \;  \omega_{\rm R}^2 \frac{d \rho_{\rm 0}}{dr}, \nonumber  \\
 && \nonumber \\
\omega^2 & = & ( \omega_{\rm R} + i \gamma)^2  \approx \omega_{\rm R} ^2 + 2 i  \gamma  \omega_{\rm R}, 
\nonumber \\
&& \nonumber \\ 
\omega^2 - \omega_{\rm A}^2 & \approx & 2 i \gamma \omega_{\rm R}, \nonumber  \\
&& \nonumber \\
\left \{\rho_{\rm 0} (\omega^2 - \omega_{\rm A}^2) \right \} ^2
& \approx  &   - 4 \,\omega_{\rm R}^4 \,(\rho_{\rm 0}(r_{\rm A}))^2 \; 
\left(\frac{\mid \gamma \mid}{\omega_R} \right) ^2,  \nonumber \\
&& \nonumber \\
-  i \frac{m}{r} \frac{\displaystyle \omega^2} {\displaystyle \left \{\rho_{\rm 0} (\omega^2 - \omega_{\rm A}^2) \right \} ^2}
\frac{\displaystyle d \rho_{\rm 0}}{\displaystyle dr}  & \approx  & 
 i \frac{m}{r_{\rm A}}\; \frac{1}{ \rho_0(r_{\rm A})\; \omega_{\rm R}^2} \; \nonumber\\
 &\times&\frac{1}{\rho_0(r_{\rm A})}\frac{d \rho_{\rm 0}}{dr} \; \pi ^2\; 
  \left\{ \frac{ \tau_{\rm D} }{\mbox{Period}}\right\}^2.
\label{IResults}
\end{eqnarray}

With the use of Eq.~(\ref{IResults}), we obtain  for the parallel component of vorticity

\begin{equation}
(\nabla \times \xi)_{\rm z}  \approx    i \frac{m}{r_{\rm A}}\; \frac{ \pi^2}{k_{\rm z} ^2} \; \frac{1}{\rho_{\rm 0}(r_{\rm A})}\frac{d \rho_{\rm 0}}{dr} 
\;\frac{1}{B_{\rm 0}^2 /\mu_{\rm 0}} \; \left\{ \frac{  \tau_{\rm D} }{ \mbox{Period}}\right\}^2  \;P'. 
\label{PVorticity1}
\end{equation}

The aim is to relate vorticity to compression. In particular, we are interested in parallel vorticity and compression. The reason  is that for a uniform plasma of infinite extent, these two quantities are mutually exclusive and characterise Alfv\'en waves and fast magneto-acoustic waves,  respectively. \cite{goossens19} have already pointed out that all wave variables are non-zero in a non-uniform plasma so that there are  no pure magneto-acoustic waves and no pure Alfv\'en waves.  

For a pressure-less plasma with a straight magnetic field, the expression for $\nabla \cdot \vec{\xi} $ is  simply
$\nabla \cdot \vec{\xi} =  - P'  /  (B_{\rm 0}^2 /\mu_{\rm 0})$.    Hence at $r = r_{\rm A}$

\begin{equation}
\frac{ \mid (\nabla \times \xi)_{\rm z} \mid} {\mid \nabla \cdot \vec{\xi} \mid} ]_{r_{\rm A}}  \approx
\frac{\mid m \mid }{r_{\rm A}}\; \frac{ \pi^2}{k_{\rm z} ^2} \; \frac{1}{\rho_0(r_{\rm A})}
\mid \frac{d \rho_{\rm 0}}{dr} \mid \; \left\{ \frac{\tau_{\rm D} }{ \mbox{Period}}\right\}^2 . 
\label{PVorticity2}
\end{equation}

So far, we have not made any assumption about the variation in the equilibrium density in the non-uniform layer or about the thickness of the non-uniform layer.  We try to obtain an estimate for 

$$
\frac{1}{\rho_0(r_A)} \mid \frac{d \rho_0}{dr} \mid 
.$$
We assume that the variation in density is confined to a non-uniform layer of thickness $l$ and has a steepness $\alpha,$ 

\begin{eqnarray*}
r_{\rm A} & \approx &  R , \;\; \rho_{\rm 0}(r_{\rm A}) \approx  \frac{\rho_{\rm i} + \rho_{\rm e}}{2}. \\\\
\mid \frac{d \rho_{\rm 0}}{dr} \mid & \approx & \alpha \; \frac{\rho_{\rm i} - \rho_{\rm e}}{l/R} \frac{1}{R},  \;\;
\frac{1}{\rho_0(r_{\rm A})} \mid \frac{d \rho_{\rm 0}}{dr} \mid  \approx  2 \;\frac{\rho_{\rm i} - \rho_{\rm e}}{\rho_{\rm i} + \rho_{\rm e}}\;\frac{\alpha}{l/R} \frac{1}{R}. 
\end{eqnarray*}
Hence 

\begin{equation}
\frac{ \mid (\nabla \times \xi)_{\rm z} \mid} {\mid \nabla \cdot \vec{\xi} \mid} ]_{r_{\rm A}}  \approx
\mid m \mid  \; \frac{ \pi^2}{(k_{\rm z} R) ^2} \; 2  \;\frac{\rho_{\rm i} - \rho_{\rm e}}{\rho_{\rm i} + \rho_{\rm e}}\;\frac{\alpha}{l/R} 
\; \left\{ \frac{ \tau_{\rm D}}{ \mbox{Period}}\right\}^2. 
\label{PVorticity3}
\end{equation}

This expression gives the dependence of 
$ \mid (\nabla \times \xi)_{\rm z} \mid $ on $k_{\rm z} R$, $l/R$, and density contrast for a  variation in density with steepness $\alpha$.  Other prescriptions  for the variation of density can be used. We note that there is a hidden dependence on $k_{\rm z} R$, $l/R$, and density contrast because $\tau_{\rm D}/\mbox{Period}$ also depends on these quantities. Nevertheless, this  simple formula is relatively good for the interpretation of numerical results obtained, for example, by  \cite{goossens12}. We improve this if we cheat slightly.  In the TT and TB approximation, analytic expressions exist for $\tau_{\rm D} / \mbox{Period}$.  With the use of expression  (\ref{TauAG}) for  $\tau_{\rm D} / \mbox{Period,}$ we can rewrite Eq.~(\ref{PVorticity3}) as 

\begin{equation}
\frac{\mid (\nabla \times \vec{\xi})_{\rm z} \mid }  {\mid \nabla \cdot \vec{\xi}  \mid} ]_{r_{\rm A}}  \approx \frac{1}{\mid m \mid}
\; \frac{32} {\pi^2}  \; \frac{1}{(k_{\rm z} R) ^2}\;\frac{\rho_{\rm i} + \rho_{\rm e}}{\rho_{\rm i} - \rho_{\rm e}} \;\frac{\alpha^3}{(l/R)^3}.
\label{PVorticity4}
\end{equation}

We  now concentrate on the transverse components  $(\nabla \times \xi)_{\rm r} $ and  $(\nabla \times \xi)_{\varphi}$. Expressions for these two components are given in the first two equations of expression 
(\ref{EqStrFieldB}). 

These two quantities are non-zero everywhere. They do not require non-uniformity to be non-zero. We now try to determine their values at and close to the position $r$= $r_{\rm A}$ where $\omega$ = $\omega_{\rm A}(r_{\rm A})$. We recall that at this position,

$$
\rho_{\rm 0} (\omega^2 - \omega_{\rm A}^2 )   \approx 2 \;  i \;  \frac{\gamma}{\omega_{\rm R}}    \;\rho_0(r_{\rm A}) \; \omega_{\rm R,} ^2
$$
so that for the radial component 

\begin{eqnarray}
(\nabla \times \xi)_{\rm r}   & \approx & -i \;  \frac{m}{k_{\rm z} \;r_{\rm A}}   \pi \;\frac{\tau_{\rm D}}{\mbox{Period}}\; \frac{1}{\; B_{\rm 0}^2 / \mu_{\rm 0}}\; P' \nonumber \\ \nonumber\\
& \approx & i \frac{m}{k_{\rm z} \;r_{\rm A}}\; \pi \; \frac{\tau_{\rm D}}{\mbox{Period}}  \;\nabla \cdot \vec{\xi}.
\label{RVorticity1}
\end{eqnarray}
Hence

\begin{equation}
\frac{ \mid (\nabla \times \xi)_{\rm r} \mid} {\mid \nabla \cdot \vec{\xi} \mid} ]_{r_{\rm A}}  
\approx \frac{\mid m\mid  \pi}{k_{\rm z} R}\; \frac{\tau_{\rm D}}{\mbox{Period}} ,
\label{RVorticity2}
\end{equation}
where we have taken $r_{\rm A}  \approx   R $. 

With the approximate expression~(\ref{TauAG}) for  $\tau_{\rm D} / \mbox{Period}$  for 
a variation in density of steepness $\alpha,$ this can be rewritten as

\begin{equation}
\frac{ \mid (\nabla \times \xi)_{\rm r} \mid} {\mid \nabla \cdot \vec{\xi} \mid} ]_{r_{\rm A}}  
\approx \frac{1}{k_{\rm z} R}  \;\frac{4}{\pi}\; \;\frac{\rho_{\rm i} + \rho_{\rm e}}{\rho_{\rm i} -  \rho_{\rm e}}\;\frac{\alpha}{(l/R)} .
\label{RVorticity3}
\end{equation}

The azimuthal component  of vorticity can be written as

\begin{equation}
(\nabla \times \xi)_{\varphi}   \approx \frac{1}{\pi \;  k_{\rm z}}\; \; \frac{\tau_{\rm D}}{\mbox{Period}}
\frac{d P'}{dr} \frac{1} {B_{\rm 0}^2 / \mu_0} \\
\label{AzVorticity1}
\end{equation}
and 

\begin{equation}
\frac{ \mid (\nabla \times \xi)_{\varphi} \mid} {\mid \nabla \cdot \vec{\xi} \mid} ]_{r_{\rm A}}  
 \approx  \frac{1}{\pi \;  k_{\rm z}}\; \; \frac{\tau_{\rm D}}{\mbox{Period}} \;
\frac{1}{P'} \; \mid \frac{d P'}{dr}\mid.
\label{AzVorticity2}
\end{equation}

\begin{figure}[]
\centering
\includegraphics[width=0.99\columnwidth]{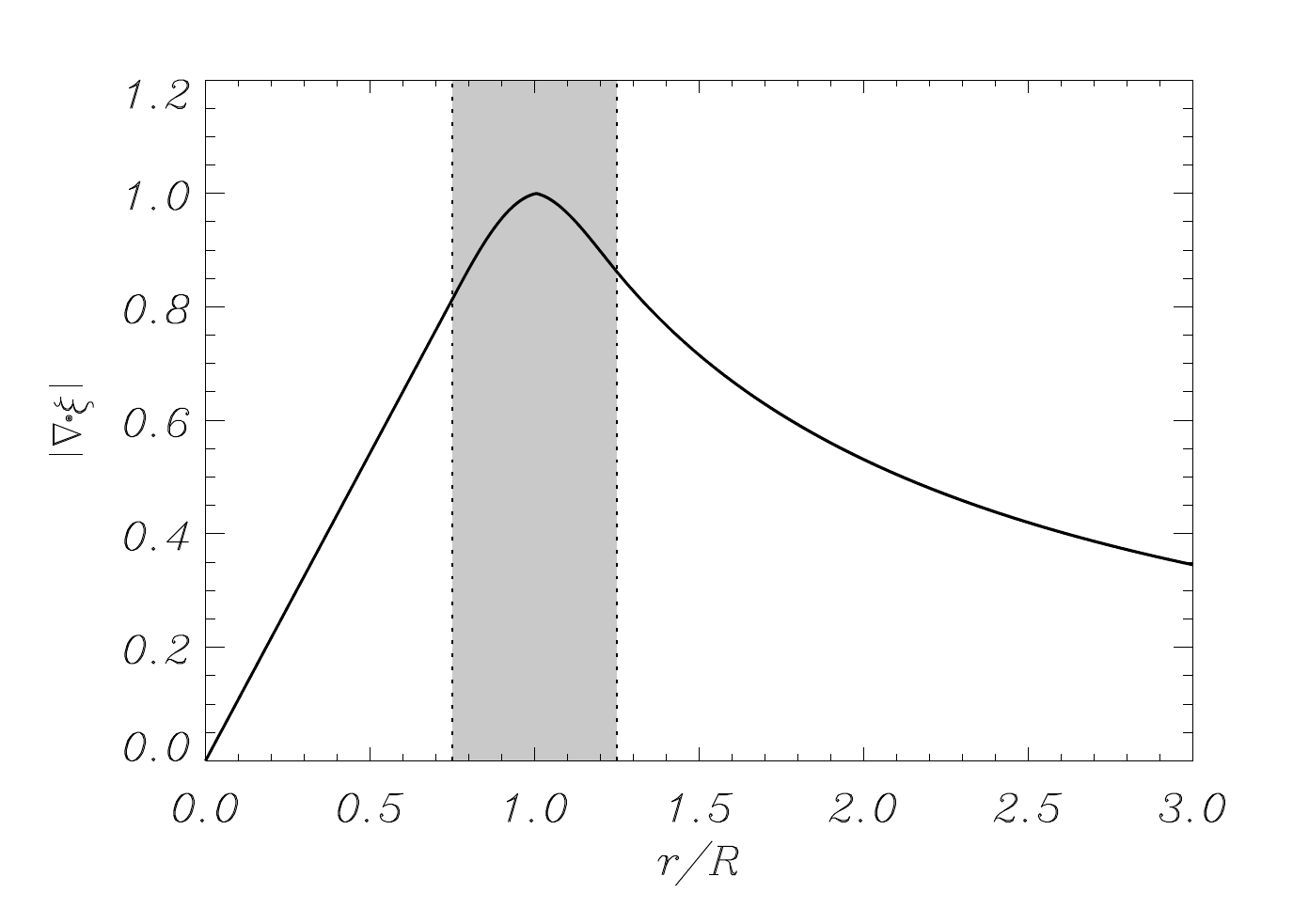}
\caption{Absolute value of the compression as a function of the radial position in a flux tube with $l/R= 0.5$, $k_{\rm z} R = 0.1$, and  $\rho_{\rm i}/\rho_{\rm e} = 3$.  The normalisation $\mbox{max}\left\{\mid \nabla \cdot \xi \mid \right \} = 1$ has been used. The shaded zone denotes the non-uniform region.}
\label{fig:f1}
\end{figure}

For a straight field the Eulerian perturbation of total pressure is a conserved quantity \citep[see e.g.][]{sakurai91,tirry96}. In non-stationary ideal MHD, it is approximately conserved when the damping is weak \citep{soler13}
and does not undergo strong spatial variations.  This is also verified by numerical computations. This  leads us  to use  $ \frac{d P'}{dr} \approx P'/R$ as an  estimate for  the derivative.  In Fig.~\ref{fig:f1} the compression, which is proportional to  $P'$, is plotted as  a function of the radial position  for the fundamental radial mode. This figure shows that a linear function is a very good representation for $P'$.  With this approximation we obtain

\begin{equation}
\frac{ \mid (\nabla \times \xi)_{\varphi} \mid} {\mid \nabla \cdot \vec{\xi} \mid} ]_{r_{\rm A}}  
 \approx  \frac{1}{\pi \;  k_{\rm z}  R }\; \; \frac{\tau_{\rm D}}{\mbox{Period}}.
\label{AzVorticity4}
\end{equation}

With the use of expression  (\ref{TauAG}) for  $\tau_{\rm D} / \mbox{Period,}$ we can rewrite Eq.~(\ref{AzVorticity4}) as

\begin{equation}
\frac{ \mid (\nabla \times \xi)_{\varphi} \mid} {\mid \nabla \cdot \vec{\xi} \mid} ]_{r_{\rm A}}  
\approx  \frac{1}{\mid m \mid} \frac{1}{ k_{\rm z}  R }\; \; \frac{4}{\pi ^3} \; 
\frac{\rho_{\rm i} + \rho_{\rm e}}{\rho_{\rm i} -  \rho_{\rm e}}\;\frac{\alpha}{(l/R)} .
\label{AzVorticity5}
\end{equation}

We recall that these expressions only apply in the non-uniform part of the loop, and strictly speaking, they apply only close to the position $r_{\rm A}$ where $\omega = \omega_{\rm A}(r_{\rm A})$. These expressions give us a good description for understanding what happens with the spatial solutions when a wave undergoes resonant damping.

\section{Spatial variation for resonant absorption for a coronal loop}

Here we apply our results from the previous two sections to understand and predict the spatial  behaviour of MHD waves that undergo resonant absorption in the equilibrium coronal loop defined in Section 3.  The aim here is to obtain a simple understanding of how the fundamental quantities compression and vorticity are influenced by the non-stationary behaviour of the resonantly damped wave.   

Compression and transverse vorticity do not require non-uniformity. They are non-zero everywhere in the plasma column. However, parallel vorticity is zero in a uniform plasma. It requires a spatial variation of the equilibrium magnetic field and/or the equilibrium density. For a constant axial magnetic field, the spatial variation in density causes the non-zero parallel vorticity. Hence we arrive at the following situation for the resonant damping of transverse waves in the Alfv\'{e}n continuum in the equilibrium model of Section 3:   In the uniform internal part of the loop,  $ 0 \leq r \leq R - l/2, $ and in the uniform external part of the loop, $ R + l/2  \leq  r < \infty,$ we expect compression and  transverse vorticity. We do not expect parallel vorticity. In the non-uniform part of the loop, we expect compression and both transverse and parallel vorticity. In the vicinity of the resonant point  $ r = r_{\rm A} $ , the parallel vorticity is expected to dominate  the transverse vorticity,

\begin{equation}
\mid( \nabla \times \vec{\xi})_z  \mid \gg  \mid( \nabla \times \vec{\xi})_r \mid \approx
 \mid( \nabla \times \vec{\xi})_{\varphi} \mid,
\label{Bigz}
\end{equation}
so that the MHD wave is almost a pure Alfv\'{e}n wave. 

In order to verify these predictions based on approximate analytic theory, now we consider the semi-analytic approach of \cite{soler13} in  non-stationary ideal MHD. The method was inspired by an early work by \cite{hollweg90b} in a simplified Cartesian configuration. We give a brief summary of the technique. We express the perturbations in the non-uniform part of the loop as a Frobenius series that includes a singular term accounting for the effect of the Alfv\'en resonance. This series expansion allows us to connect through the non-uniform layer the perturbations in the internal medium to those in the external medium. Thus, we obtain a dispersion relation for all trapped modes with any value of the azimuthal wavenumber, $m$. However, we focus on $m=1$. The dispersion relation is valid for arbitrary values of $l/R$. When the nonuniform layer is thin, that is, $l/R\ll 1$, the obtained dispersion relation consistently provides the same results as those of the TTTB approximation. The numerical part of the method consists of numerically solving the transcendental dispersion relation to obtain the complex frequency, $\omega$, for a fixed $k_{\rm z}$.  When the frequency is known, we can compute the spatial behaviour of the perturbations. We refer to \cite{soler13} for a more detailed explanation of the method.

\begin{figure*}[!h]
\centering
\includegraphics[width=0.99\textwidth]{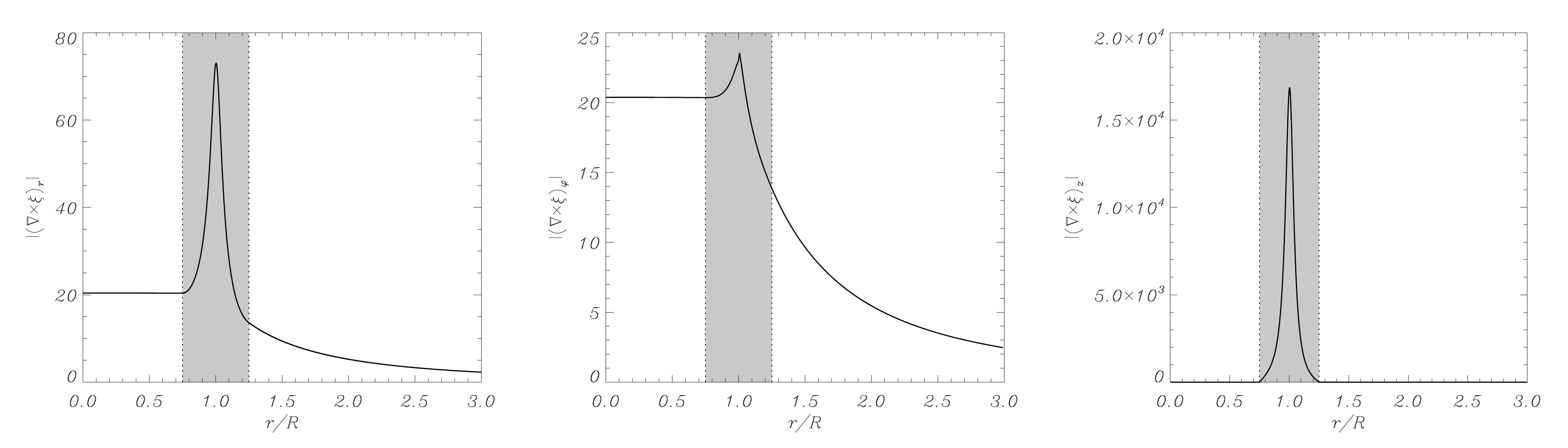}
\caption{Absolute value of the radial (left), azimuthal (centre), and parallel (right) vorticity components as functions of the radial position in the same flux tube as in Fig.~\ref{fig:f1}.}
\label{fig:f2}
\end{figure*}

\begin{figure*}[!h]
\centering
\includegraphics[width=0.95\textwidth]{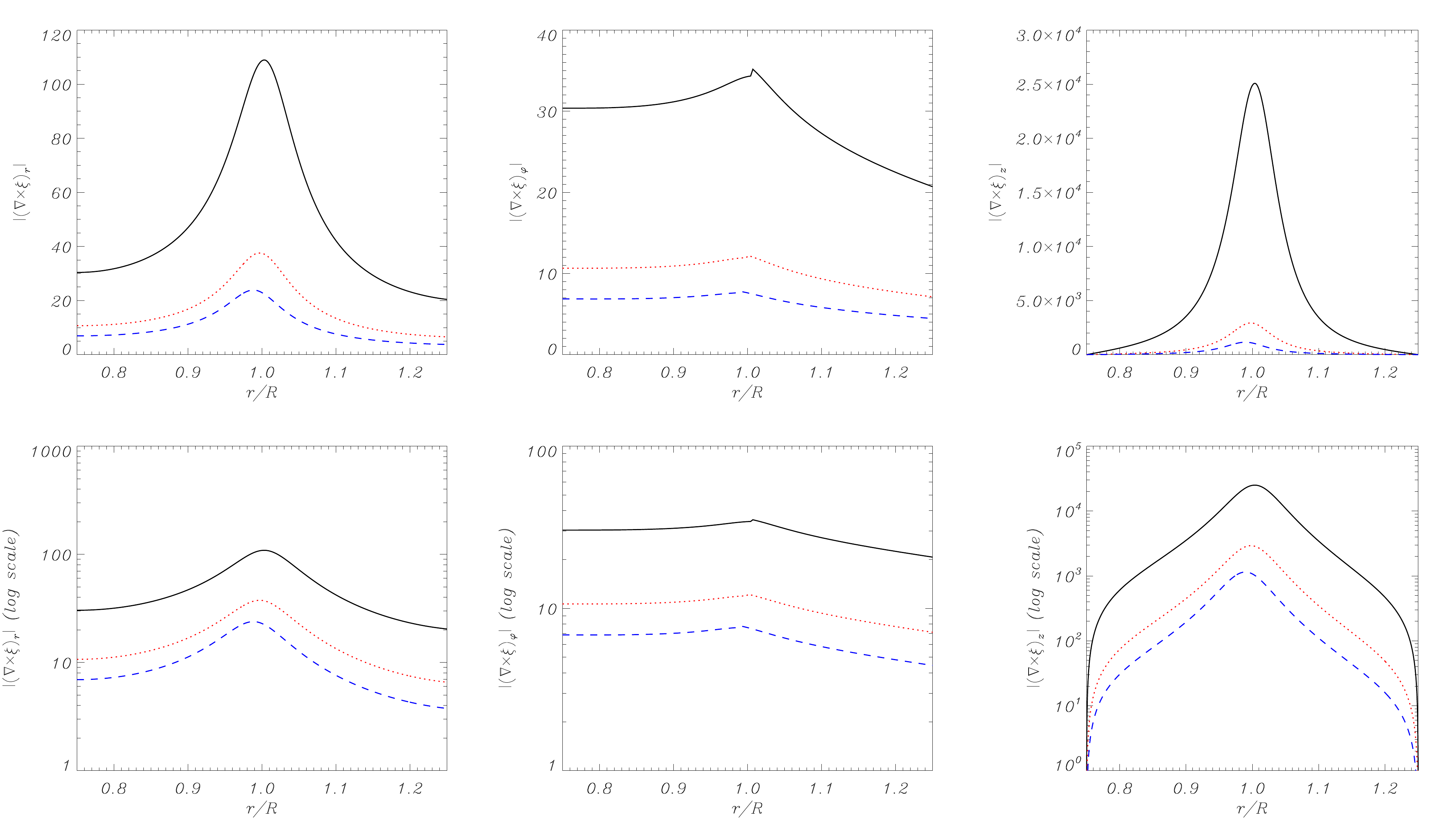}
\caption{Absolute value of the radial (left), azimuthal (centre), and parallel (right) vorticity components in the non-uniform part of a flux tube with $l/R=0.5$ and $\rho_{\rm i}/\rho_{\rm e} = 2$. The top panels are in linear scale, and the bottom panels are in logarithmic scale. The different line styles denote $k_{\rm z}R = 0.1$ (solid black line), $k_{\rm z}R=0.3$ (dotted red line), and $k_{\rm z}R=0.5$ (dashed blue line).}
\label{fig:f3}
\end{figure*}

\begin{figure*}[!h]
\centering
\includegraphics[width=0.95\textwidth]{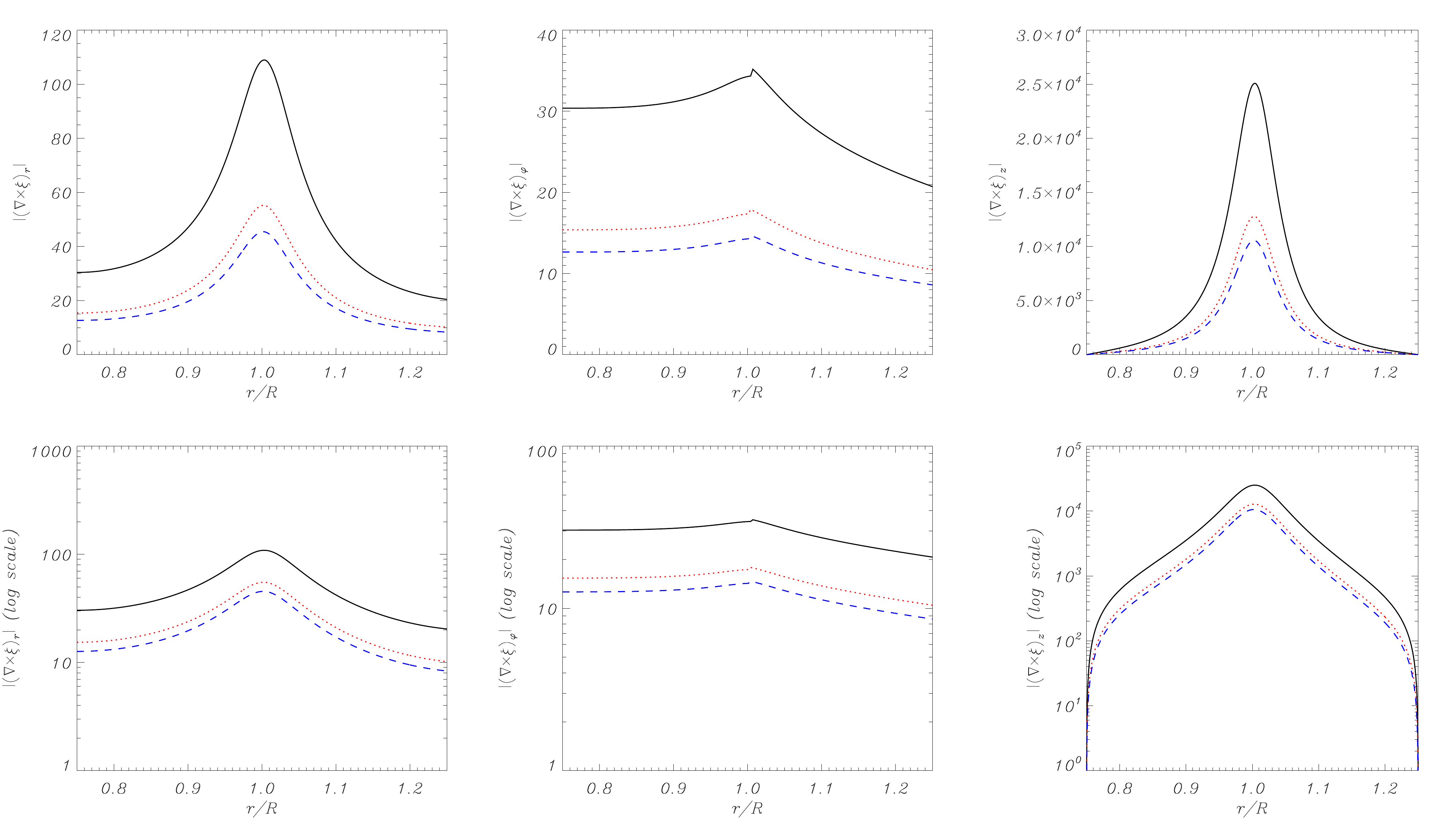}
\caption{Same as Fig.~\ref{fig:f3}, but with $k_{\rm z}R = 0.1$ and different values of the density contrast: $\rho_{\rm i}/\rho_{\rm e}=2$ (solid black line), $\rho_{\rm i}/\rho_{\rm e}=5$ (dotted red line), and $\rho_{\rm i}/\rho_{\rm e}=10$ (dashed blue line).}
\label{fig:f4}
\end{figure*}

\begin{figure}[!t]
\centering
\includegraphics[width=0.99\columnwidth]{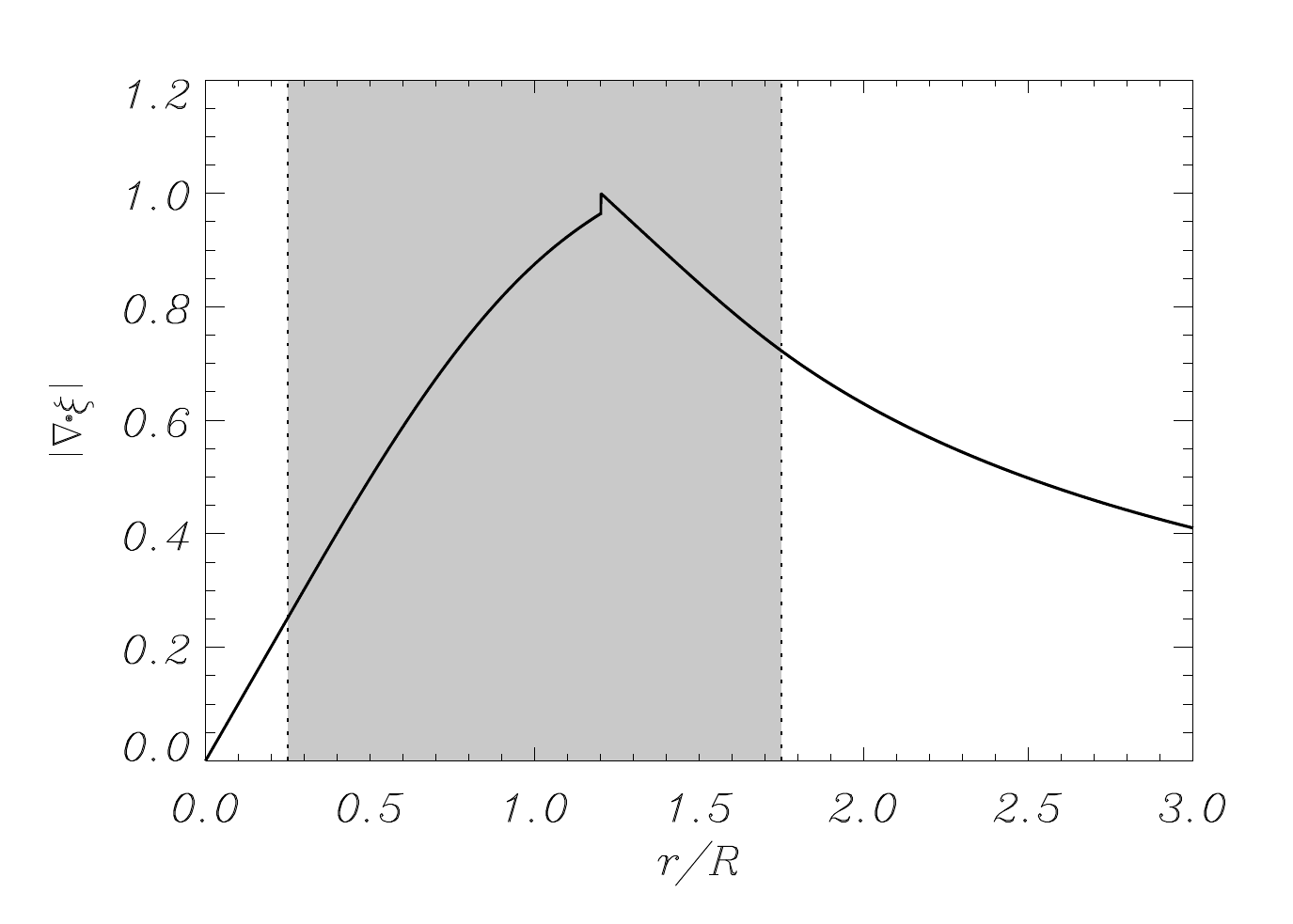}
\caption{Same as Fig.~\ref{fig:f1}, but with $l/R= 1.5$.}
\label{fig:f5}
\end{figure}

\begin{figure*}[!h]
\centering
\includegraphics[width=0.99\textwidth]{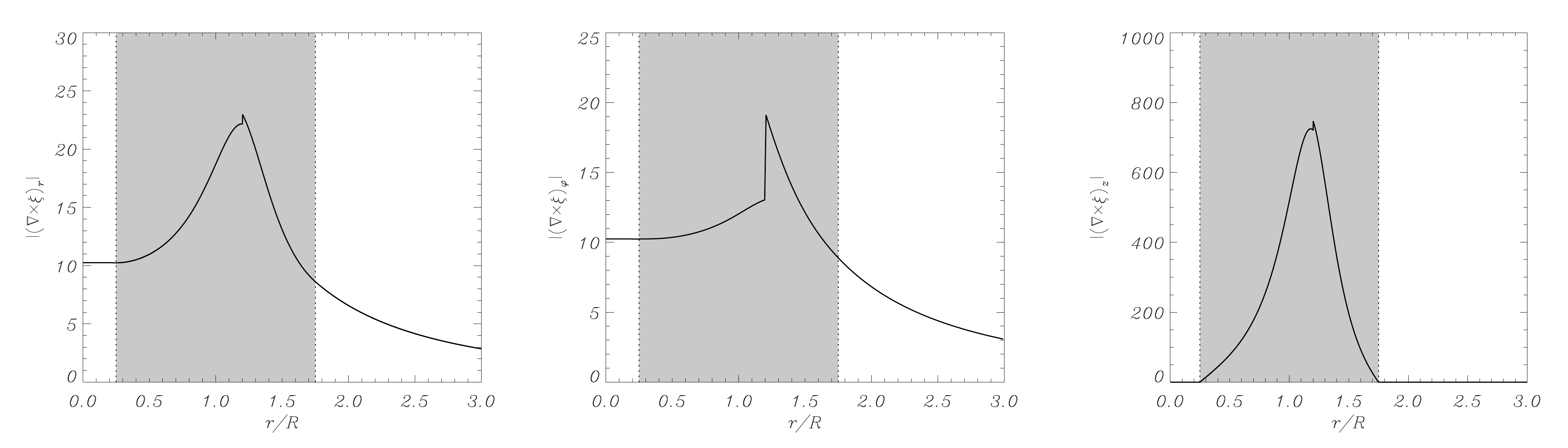}
\caption{Same as Fig.~\ref{fig:f2}, but with $l/R= 1.5$.}
\label{fig:f6}
\end{figure*}

\cite{soler13} studied the components of Lagrangian displacement $\xi_r$, $\xi_{\varphi,}$ and the Eulerian perturbation of total pressure, $P'$. Compression and the components of vorticity have now been computed and are presented in Figs.~\ref{fig:f1}-\ref{fig:f6}. A similar exercise was made for  the  parallel vorticity  in non-stationary resistive MHD  by \cite{goossens12} (see their figure 4).  The normalisation $\mbox{max}\left\{\mid \nabla \cdot \xi \mid \right \} = 1$ was used in all figures.  A sinusoidal variation for the density in the nonuniform part was assumed.

In Figs.~\ref{fig:f1} and \ref{fig:f2}  we plot compression and the components of vorticity on the interval $r/R\in[0, 3]$ for a loop with a relatively thin non-uniform layer with $l/R= 0.5$.  We also used $k_{\rm z} R = 0.1$ and  $\rho_{\rm i}/\rho_{\rm e} = 3$. This case can be directly compared with the approximate TTTB results obtained in the previous sections. Figure~\ref{fig:f1} shows that nothing particular happens for compression.  The behaviour does not substantially differ from that found in the stepwise constant case. Figure~\ref{fig:f2} shows that the transverse components of vorticity are non-zero in the whole domain; the parallel component of vorticity is only non-zero in the non-uniform part of the loop. The parallel component of vorticity is far larger than the transverse components. In Fig.~\ref{fig:f2} the maximum value of the parallel component is at least four orders of magnitude higher than the maximum values of the transverse components.  In turn, the amplitudes of the two transverse components are of the same order of magnitude. All these results agree with the analytic predictions in the TTTB approximation.

Next we consider $l/R= 0.5$ as before and compute the components of vorticity for various values of $k_{\rm z} R$ and $\rho_{\rm i}/\rho_{\rm e}$. In Fig.~\ref{fig:f3} we considered three values of $k_{\rm z} R$ = 0.1, 0.3, and 0.5 and $\rho_{\rm i}/\rho_{\rm e}$ = 2, while in Fig. ~\ref{fig:f4} we considered three values of $\rho_{\rm i}/\rho_{\rm e}$ =2, 5, and 10 and $k_{\rm z} R$=0.1. Because the parallel vorticity is only non-zero in the non-uniform transitional layer, the components are plotted on the interval $r\in[ R - l/2, R + l/2]$ alone. The results are plotted in both linear and logarithmic scales for better visualisation, and we recall that $\mbox{max}\left\{\mid \nabla \cdot \xi \mid \right \} = 1$ in all cases. The spatial profile of vorticity is largely unaffected when either $k_{\rm z}R$ or $\rho_{\rm i}/\rho_{\rm e}$ are modified. This is so because the spatial behaviour of vorticity is largely determined by the spatial variation of density, which is the same in all cases. However, the relative amplitudes of the three components of vorticity depend upon the considered $k_{\rm z} R$ and $\rho_{\rm i}/\rho_{\rm e}$. The larger $k_{\rm z} R$, the lower  the vorticity. The amplitude of the field-aligned component of vorticity decreases with $k_{\rm z}R$ as $\left(k_{\rm z} R\right)^{-2}$, approximately, as Eq.~(\ref{PVorticity3}) predicts, while both radial and azimuthal components behave as  $\left(k_{\rm z} R\right)^{-1}$ according to Eq.~(\ref{RVorticity2}) and (\ref{AzVorticity4}). Equivalently, the higher the density contrast, $\rho_{\rm i}/\rho_{\rm e}$, the lower the vorticity. The dependence on density contrast is also consistent with the analytic TTTB formulas, although there is no such clear dependence as in the case of $k_{\rm z} R$. 

We have confirmed so far that the analytic predictions in the TTTB approximation are consistent with the full solution provided by the method of \cite{soler13} when the non-uniform transition is thin. Now, we can be more ambitious and fully exploit the method by computing results  beyond the range of applicability of the approximations. Figures~\ref{fig:f5} and \ref{fig:f6}  display compression and the components of vorticity, respectively, for a loop with a thick non-uniform layer of $l/R= 1.5$. The other parameters are the same as in Figs.~\ref{fig:f1} and \ref{fig:f2}.

When we compare the plots of compression for a thin (Fig.~\ref{fig:f1}) and a thick (Fig.~\ref{fig:f5}) non-uniform transition, a distinct feature is obvious. In the case of a thick transition, compression displays a small jump at the resonance position. In order to understand the behaviour of compression, we recall that compression is proportional to $P'$ and we resort to the Frobenius series of $P'$. In general, the full solution provided by the method of Frobenius is difficult to handle \citep[see][]{soler13}. To illustrate the present discussion, it suffices to consider the first non-zero terms in the Frobenius series of $P'$ in the nonuniform layer, 

\begin{equation}
P' \approx S_0  +   \left( A_0 + S_0 \frac{m^2}{2r_{\rm A}^2}\ln \left( r - r_{\rm A}\right) \right)\left( r - r_{\rm A}\right)^2, \label{eq:pfrobenius}
\end{equation}
where $A_0$ and $S_0$ are constants. The neglected terms in Eq.~(\ref{eq:pfrobenius}) are of order $\left( r - r_{\rm A}\right)^3$ and higher. We note that in non-stationary ideal MHD, $r_{\rm A}$ as defined by the resonant condition $\omega = \omega_{\rm A}(r_{\rm A})$ is a complex quantity because $\omega$ is a complex quantity. As a consequence of this, there is not an actual resonance  but a quasi-resonance at the radial position where $r=\operatorname{Re}(r_{\rm A})$. At the quasi-resonance position, $r-r_{\rm A} = -i\,\operatorname{Im}(r_{\rm A}) \ne 0$. The imaginary part of $r_{\rm A}$ owes its existence to the non-zero damping rate, $\gamma$.    Equation~(\ref{eq:pfrobenius}) has a term proportional to $\left( r - r_{\rm A}\right)^2\ln \left( r - r_{\rm A}\right)$. This term  does not vanish at $r=\operatorname{Re}(r_{\rm A})$ as it would do at $r=r_{\rm A}$ if $r_{\rm A}$ were real. Instead, the logarithm jumps when $r-r_{\rm A}$ crosses the imaginary axis.  In the case of a thin non-uniform layer, $\gamma$ is small, so that ${\rm Im}(r_{\rm A})$ is small and can be ignored. Then, $P'\approx$~constant as in stationary ideal MHD and as the thin boundary approximation assumes \citep[see][]{hollwegyang88}. Conversely, when the nonuniform layer is thick the wave is strongly damped, non-stationarity is important, $\gamma$ is not small, and $\operatorname{Im}(r_{\rm A})$ is not negligible. Hence, the jump in $P'$. This explains the presence of the jump in compression seen in Fig.~\ref{fig:f5}. Of course, if a dissipative process is taken into account, the jumps obtained in non-stationary ideal MHD are replaced by smooth variations in non-stationary dissipative MHD  \citep[see e.g.,][]{mok85, ruderman95,wright96,vanlommel02,soler12,soler13}.

On the other hand, by comparing Figs.~\ref{fig:f2} and \ref{fig:f6}, we see that the components of vorticity have a smaller amplitude in the case of a thick transition. Remarkably, the amplitude of the field-aligned component is more than an order of magnitude smaller. The analytic TTTB approximations, although not strictly valid now,  are helpful to understand this result. According to Eq.~(\ref{PVorticity4}), the parallel component should behave as $\left(l/R   \right)^{-3}$, while both radial and azimuthal components should behave as $\left(l/R\right)^{-1}$ according to Eqs.~(\ref{RVorticity3}) and (\ref{AzVorticity5}). The predicted behaviour in the TTTB approximation explains the decrease in amplitude of the vorticity components when $l/R$ increases and why the parallel component decreases more than the perpendicular components. Finally, we also note that in Fig.~\ref{fig:f6} all the components of vorticity jump at the resonance position, whereas the jumps were not apparent in Fig. 2.

\section{Conclusions}

We used linear non-stationary ideal MHD to investigate the spatial behaviour of compression and vorticity for MHD waves that undergo resonant absorption in the Alfv\'en continuum. In linear MHD there is no interaction between waves, and the behaviour that we discussed is associated with a single MHD wave that lives in the whole space. Pure Alfv\'en waves and pure magneto-acoustic waves exist in a uniform plasma of infinite extent. In a non-uniform plasma, the MHD waves  combine the properties of the classic Alfv\'en waves and of the magneto-acoustic waves in a uniform plasma of infinite extent.  In a non-uniform plasma,  MHD waves  propagate both compression and parallel and transverse vorticity. The properties of the MHD wave depend on the properties of the background plasma.  As an MHD wave propagates through a non-uniform plasma, its properties therefore change.  When an MHD wave moves from a uniform into a non-uniform plasma, it is transformed from a fast magneto-acoustic wave into a mixed fast - Alfv\'en  wave. 

Resonant absorption is a clear example of the phenomenon that the properties of an MHD wave change when it travels through an inhomogeneous plasma. In the case of resonant Alfv\'{e}n waves, the MHD wave eventually arrives at  a position  where it  behaves as an almost pure Alfv\'{e}n wave, but with the unfamiliar property that it has pressure variations.  The total pressure perturbation and compression are non-zero everywhere. The pressure variations are essential for resonant absorption because the amount of absorbed energy and the damping rate are directly related to the pressure variations \cite[see e.g.][]{thompson93,tirry95,andries01,goossens02b,goossens08b,goossens11,arregui11}. 

Classic Alfv\'en waves are not the only waves to propagate vorticity from the photosphere to outer space.  MHD waves that undergo resonant absorption in a  non-uniform plasma can also play this role. 

\begin{acknowledgements}
M.G. was supported by the C1 grant TRACEspace of Internal Funds KU Leuven (number C14/19/089).  I.A. was supported by project PGC2018-102108-B-I00 from Ministerio de Ciencia, Innovaci\'on y Universidades and FEDER funds. R.S. acknowledges the support from grant AYA2017-85465-P (MINECO/AEI/FEDER, UE) and from the Ministerio de Econom\'{\i}a, Industria y Competitividad and the Conselleria d$'$Innovacio, Recerca i Turisme del Govern Balear (Pla de ci\`encia, tecnologia, innovaci\'o i emprenedoria 2013-2017) for the Ram\'on y Cajal grant RYC-2014-14970.
T.V.D. was supported by the  European Research Council (ERC) under the European Union's Horizon 2020 research and innovation programme (grant agreement No 724326) and the C1 grant TRACEspace of Internal Funds KU Leuven (number C14/19/089).
\end{acknowledgements}

\end{document}